# PROBLEM SOLVING
# AND THE USE OF MATH IN PHYSICS COURSES

EDWARD F. REDISH

*Department of Physics, University of Maryland*
*College Park, MD, 20742-4111 USA*

Mathematics is an essential element of physics problem solving, but experts often fail to appreciate exactly how they use it. Math may be the language of science, but math-in-physics is a distinct dialect of that language. Physicists tend to blend conceptual physics with mathematical symbolism in a way that profoundly affects the way equations are used and interpreted. Research with university physics students in classes from algebra-based introductory physics indicates that the gap between what students think they are supposed to be doing and what their instructors expect them to do can cause severe problems.

## 1. We use math differently in physics

Mathematics is commonly referred to as "the language of science" and we typically require our physics students to take mathematics as prerequisites to their study of physics. As instructors, we are often surprised by how little math our students seem to know, despite successful performances in their math classes. When students appear to have trouble with math in our physics classes, we might ask them to "study more math." But using math in science (and particularly in physics) is not just doing math. It has a different purpose – representing meaning about physical systems rather than expressing abstract relationships – and it even has a distinct semiotics – the way meaning is put into symbols – from pure mathematics.

It almost seems that the "language" of mathematics we use in physics is not the same as the one taught by mathematicians. There are many notable differences.

### 1.1. *Physicists and mathematicians label constants and variables differently*

In mathematics as typically taught, the choice of symbols tends to be narrowly restricted by category. In a one-variable-calculus class, the variable will almost always be an $x$, $y$, $z$, or $t$. Constants will typically be represented as specific numbers. If they are kept general, there will be an $a$, $b$, $c$, or $d$. In a typical calculus-I text (one-variable) not one equation in 1000 will contain more than one symbol. In physics we use many different symbols. In a typical calculus-based physics class, the equations shown in the first week have from three to six symbols or more. Of course, most of these are constants or parameters – "just numbers"[†] – specifying a connection with something physical. Equations with a single symbol are exceedingly rare — and not just in the sense of our using a larger palette of letters or our predilection to represent numbers as combinations of symbols.

---

[†] As we will see in our discussion of units, these parameters are rarely "just numbers."

- We have many different kinds of constants – numbers (2, $e$, $\pi$,...), universal dimensioned constants ($e$, $h$, $k_B$,...), problem parameters ($m$, $R$,...), and initial conditions.
- We blur the distinction between constants and variables.
- We use symbols to stand for ideas rather than quantities.
- We mix "things of physics" and "things of math" when we interpret equations.

But perhaps the most dramatic difference is the way we put meaning to our symbols.

### *1.2. Loading meaning onto symbols leads to differences in how physicists and mathematicians interpret equations*

In physics, our symbols are not arbitrary but tend to be chosen to activate a particular mental association with some physical quantity or measurement. Consider the problem shown in Fig. 1.

$$\text{If} \quad A(x,y) = K(x^2 + y^2) \quad K \text{ a constant}$$
$$\text{What is} \quad A(r,\theta) = ?$$

Fig. 1: A problem that tends to distinguish physicists from mathematicians.

I have asked this question to dozens of physicists. Almost all have a ready answer:

$$A(r,\theta) = Kr^2. \tag{1}$$

The reason is clear. Without it being specified precisely in the problem, the $x^2+y^2$ is a familiar combination. It activates coordinates in the plane and the Pythagorean theorem in the viewer's mind. The use of $r$ and $\theta$ in the second equation supports this expectation and the answer comes readily to hand using the ancillary equation $x^2 + y^2 = r^2$.[†]

A mathematician, on the other hand, would insist that the answer has to be

$$A(r,\theta) = K(r^2 + \theta^2) \tag{2}$$

since the function as defined says take the sum of the squares of the two arguments and multiply by $K$. This gives the result shown in Eq. (2). Why does this seem so wrong? You can't add $r^2$ and $\theta^2$! They have the wrong units! Of course a mathematician would have no reason to assume any of the quantities had units.

Our mathematician would argue that if you mean to change the functional form of $A$, you should use a new symbol; perhaps write

$$A(x,y) = B(r,\theta). \tag{3}$$

A physicist would still be uncomfortable with this. I can just hear her say: "But $A$ represents the vector potential. I can't use $B$ to stand for the vector potential! That would be confusing it with the magnetic field!"

---

[†] I am grateful to Corinne Manogue for introducing me to this wonderful example.

This is revealing. It implies that physicists use their idea of what physical quantity a symbol represents to decide how the math should be interpreted. Also, since physicists do indeed pay attention to functional dependence in some contexts (e.g., Lagrangians vs. Hamiltonians or thermodynamic potentials), the fact that they don't in this example implies a contextual sensitivity to how the math is used. This is not really a surprise. All languages have context dependences. Consider how you easily know whether a speaker is saying "there", "their", or "they're" from context without hesitation. Math also has such context dependences. In Eq. (2), for example, consider the two different meanings of the parentheses, "(…)", on the two sides of the equation. Neither mathematician nor physicist has trouble seeing those parentheses in different, context dependent, ways. (Sometimes, novice students do indeed confuse these two kinds of parentheses, to their instructor's surprise and dismay.)

### *1.3. Loading physical meaning onto symbols does work for us*

The fact that physicists "load" physical meaning onto symbols in a way that mathematicians do not is both powerful and useful. It allows us to work with complex mathematical quantities without introducing the fancy math that would be required to handle some issues with mathematical rigor.

*Example: Units*

One of the issues in our $A(x,y)$ example above had to do with the "units" of $x$, $y$, $r$, and $\theta$. We said that $r$ and $\theta$ had different units. What does it mean for a quantity to have "units"?

In introducing the concept of units to my introductory classes, I tell them that a quantity has a unit when we have an operational definition for assigning a number to it and that the number that results depends on the choice of an arbitrary standard. Since the standard is arbitrary, we may only equate quantities (or add them) if they change in the same way when we change our standard. Otherwise, a numerical equality that we obtain might be true for one choice of a standard but not for another. An equation that has physical validity ought to retain its correctness independent of our arbitrary choices.

Many physicists will recognize in the above the "scent of Einstein" – the idea that "a difference that makes no difference should make no difference." Einstein, Poincaré, and others around the turn of the last century introduced the idea of analyzing how physical measurements change when you change your perspective on them – rotate your coordinates (in the case of identifying vectors and tensors), hop onto a uniformly moving frame (in the case of special relativity), or make a general non-linear coordinate transformation (in the case of general relativity).

The identification with something as apparently trivial as the unit check with something as sophisticated as general relativity may seem inappropriate. But this is only because we have physical experiences about measurements that make sense to us. We can substitute our physical intuition for the statement that our equations must be covariant under the transformation of the product of three scaling groups. We can, for example, build on our intuition with simple counting to understand the difference in the way a distance and an area change when we change our standard of length measurement.

*Example: Position vs. velocity vector*

The issue of the position vs. the velocity (or displacement) vector is a second example of how we blend our sense of a physical object with a mathematical symbol in order to simplify the math and change how we interpret the symbology. Both position and velocity vectors transform in the same way under rotations of the coordinate system about the origin. But a position vector changes when the origin shifts (is an affine vector), while a velocity vector is independent of the position of the origin (is an affine scalar) since it depends on position only through the difference of two position vectors (a displacement).

We don't typically fuss about this in an introductory physics class. Even courses in group theory for physicists tend to ignore the issue of affine transformations since they are so easily handled by our physical sense of how positions behave.

### *1.4. Blending physical meaning with math changes the way we look at equations*

The blending of physical meaning with mathematical symbols not only affects how we interpret particular symbols, it affects how we view equations. Two ways that this occurs are first, through seeing equations as relations, not as calculational methods, and second, through "filtering the equation through the physics."

*Seeing the equation as representing relationships: Limiting cases*

Students in introductory physics have a strong inclination to put numbers into their equations as soon as they know them. This makes the equations look more like the equations in their math classes and makes them seem more familiar. However, doing this leads to difficulties. When students put in numbers, they tend to drop the units. They say "I'll put them in at the end since I know how they have to come out." This, of course, loses the advantages of using the units as a check on errors or inappropriate mixtures of units. But for me, the second problem is the more severe.

I believe that a primary reason we physicists prefer to keep our constants as symbols all the way to the end of a calculation rather than putting numbers in at the beginning is that we see an equation as relationships among physical measurements. It is not just a way to calculate a result. It is a way to generate a whole ensemble of results: not just the one you are currently calculating, but all possible situations with the same physics but different values for the parameters. This is a rather dramatic shift of outlook and one that we want to help our students master.

A useful class of problem for encouraging students to work with symbols is the limiting case problem. A simple example is shown in Fig. 2. Taking the limiting case of either of the two masses going to zero (or infinity) is an example of considering an ensemble of experiments rather than just a single one and is also a nice example of physicists' willingness to treat constants (the masses) as variables.

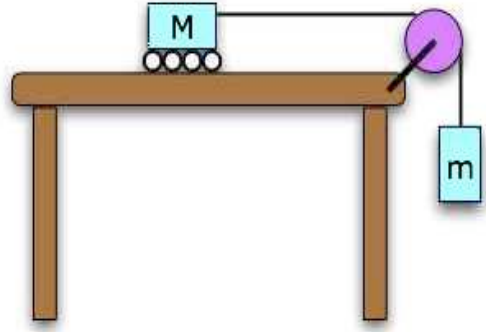

A mass M on nearly frictionless wheels is attached by a string to a mass m over a frictionless pulley. The mass M is held fixed and then released.

(a) Find the acceleration of the two masses.

(b) Find the acceleration of the two masses in the case that m → 0; in the case that M → 0. Do these results make sense?

Fig. 2: The half-Atwood's machine; a useful example for learning to take limiting cases of a problem's parameters.

*Filtering the equation through the physics: The photoelectric effect*

Another striking effect of the fact that physicists blend physical concepts with mathematical symbology is that the way an equation is used can be strongly affected. A nice example is the equation for the photoelectric effect.

I gave the problem shown in Fig. 3 to my class of third semester engineering physics students. It appears quite simple. All that I am asking them to do is to realize that a longer wavelength corresponds to a lower frequency. If the original light did not have enough energy to knock out an electron, then a longer wavelength would have even less energy – certainly not enough to knock out an electron.

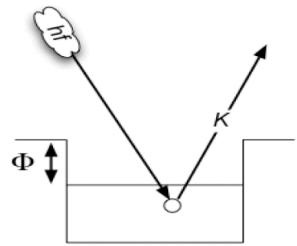

If a wavelength of light $\lambda$ leads to no electrons being emitted at a zero stopping potential, what will happen if we choose a longer wavelength?

Fig. 3: A simple problem in the photoelectric effect.

I was quite taken aback at the result. Almost one quarter of my students said something like, "We use Einstein's photoelectric equation. Changing the wavelength changes the frequency. Since we had zero before, if we change *f* we won't have zero anymore so we should get electrons out."

My students were interpreting the Einstein equation in quite a different way from the way I expected them to. The equation is

$$eV_0 = hf - \Phi \qquad (4)$$

where $V_0$ is the electrostatic potential required to stop all the electrons, $f$ is the frequency of the light, $h$ is Planck's constant, and $\Phi$ is the work function (essentially the binding energy of the electron that is the most weakly bound in the metal). A physicist sees this as a conservation of energy equation. The energy of the photon minus the binding energy of the electron in the metal is equal to the kinetic energy of the electron when it is knocked out of the metal. The electron's charge times the stopping potential is the size of the energy hill we have to create so that no electron will have enough energy to roll up it.

This conceptual view of the situation serves as a filter through which we use the equation. Since we know kinetic energy must be positive (ignoring quantum tunneling effects), we test first to see if there is enough energy in the photon to produce an electron with a positive kinetic energy. If there isn't, then we do not apply the equation.

But the conceptual idea that the $eV_0$ term represents a kinetic energy is not present in the equation itself. We bring it in by blending our physical interpretation with the math. Once we do this, we do not require the presence of the Heaviside function $\theta(hf - \Phi)$ that should really be included in Eg. (4) to correctly represent the presence of a threshold.

These examples are a clear demonstration that our use of equations in physics is a significantly more complex cognitive process than students have come to expect from their math classes. We use symbols that carry ancillary information not otherwise present in the mathematical structure of the equation. We use more complex quantities than in math class and use them tacitly. We interpret our equations through knowledge of physical systems — which adds information.

## 2. Physicists and mathematicians have different goals for the use of math

It's not just the way we read and use our equations that are different from math. Our goals are different. We don't just want to explore ways of solving equations, we want to describe, learn about, and understand physical systems.

### 2.1. A model of mathematics in science

A model describing the bare bones of how we use math in physics (and in other sciences as well) is shown in Fig. 4.

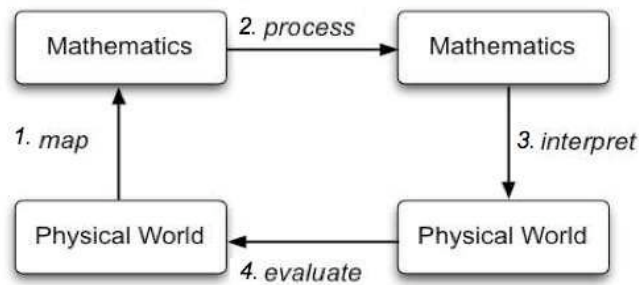

Fig. 4: A model for the use of math in science.

We begin in the lower left corner by choosing a physical system we want to describe. Within this box, we have to decide what characteristics of the system to pay attention to and what to ignore. This is a crucial step and is where much of the skill or "art" in doing physics lies. Looking at a complex physical system and deciding what are the critical elements that must be kept, what are trivial effects that can be ignored, and what are somewhat important effects that can be ignored at first and corrected later is what we mean by "getting the physics right." Einstein said it best: *Everything should be as simple as possible – but not simpler.*

Once we have decided what we need to consider, we than do step 1: map. We map our physical structures into mathematical ones – create a mathematical model. To do this, we have to understand what mathematical structures are available and what aspects of them are relevant to the physical characteristics we are trying to model.

Now that we have mathatized our system, we are ready for step 2: process. We can use the technology associated with the math structures we have chosen to transform our initial description. We may be solving an equation or deriving new ones. But once we have done that, we are a long way from finished.

We still have to do step 3: interpret. We see what our results tell us about our system in physical terms and then do step 4: evaluate. We have to evaluate whether our results adequately describe our physical system or whether we have to modify our model.

### 2.2. Our traditional instruction of math in physics may not give enough emphasis to some of the critical steps in this model

Our traditional approach does not help students focus on some of these important steps. We tend to provide our students with the model ready made, and we may be exasperated – or even irritated – if they focus on details that we know to be irrelevant. We tend to let them do the mathematical manipulations in the process step, and we rarely ask them to interpret their results and even less often ask them to evaluate whether the initial model is adequate.

At the introductory level, our exams often require only one-step recognition, giving "cues" so we don't require our students to recognize deep structures. When they don't succeed on their own with complex problem solving, we tend to "pander" by only giving simple problems. We often don't recognize what's complex in a problem for a student, and that makes it hard to design appropriate and effective problems.

These problems don't only occur at the introductory level (and there, the results of physics education research are beginning to improve these issues somewhat), but all throughout the physics major's curriculum. At the more advanced levels, we give our students more complex math to process, but we rarely give them an opportunity to explore the other aspects of the model. I have been trying to find and develop problems that address these issues. One that I use in my Intermediate Methods of Mathematical Physics class is shown in Fig. 5.

> The pair of coupled ordinary differential equations
>
> $$\frac{dx}{dt} = Ax - Bxy$$
>
> $$\frac{dy}{dt} = -Cy + Dxy$$
>
> are referred to as the Lotka-Volterra equations and are supposed to represent the dependence of the populations of a predator and its prey as a function of time. The constants *A*, *B*, *C*, and *D* are all positive.
>
> (a) Which of the variables, *x* or *y*, represents the predator? Which represents the prey? What reasons do you have for your choice?
>
> (b) What do the parameters *A, B, C,* and *D* represent? Why do you say so?
>
> (c) Do you expect that these equations include all the relevant phenomena? Or have some important effects been omitted? Explain why you think so.

Fig. 5: The Lotka-Volterra equations. A problem that stresses modeling and interpretation

### 3. Student expectations about how to do math in science can cause problems

Students' failure to blend physics and math leads them to expect to transform their problem solving in physics into problem solving in math. The way they do this can be analyzed in terms of structures of student expectations. Physics education research has demonstrated that students' expectations can play a powerful role in how they use the knowledge they have in our physics classes [1]. Student expectations also play a powerful role in how they think they are supposed to use math in their physics (or science) classes.

In a study of problem solving in algebra-based physics [2][3], Jonathan Tuminaro videotaped students working together to solve physics problems and made an interesting observation. While trying to solve a problem, students choose a local goal or subgoal and tend to work on that task within a locally coherent organizational framework — one that employs only a fraction of their problem-solving resources. They may "shift gears" to a new (very different) activity when one fails to prove effective.

We refer to such a coherent local (in time) pattern of activity for building knowledge or solving a problem as an *epistemic game* (or *e-game* for short) [4].[†] An epistemic game has a goal, moves (allowed activities), and an endstate (a way of knowing when the game has been won). The important thing about this observation is the observation that e-games are *exclusionary*; that is, they limit the moves that can be use to the ones within the game. Other resources that might be appropriate are not accessed, even if they might be useful.

---

[†] Collins and Ferguson [4] introduced this term but used it in a normative sense – to describe activities carried out by experts that need to be taught. We use the term in an ethnographic sense – as a way to describe activities that we see students do.

In watching about 50 hours of video of student groups solving problems, Tuminaro identified six commonly used e-games: Mapping Meaning to Mathematics, Mapping Mathematics to Meaning, Physical Mechanism, Pictorial Analysis, Recursive Plug-and-Chug, and Transliteration to Mathematics [5]. Flow charts describing the primary moves in two of these games are shown in Figs. 6 and 7.

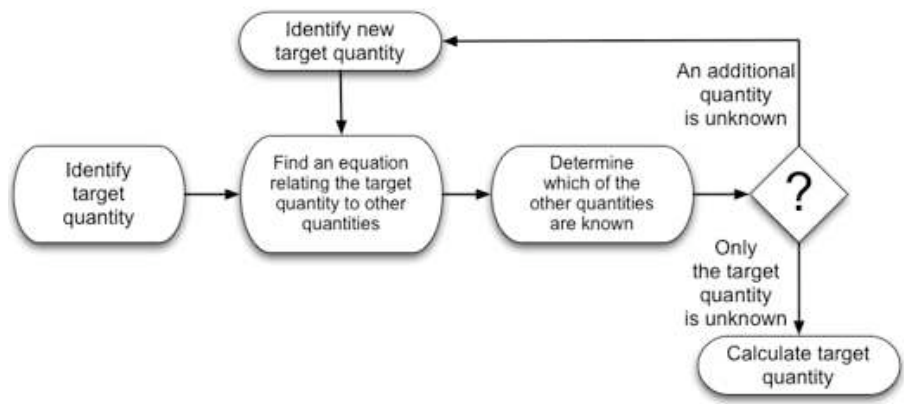

Fig. 6: Recursive plug-and-chug -- An epistemic game that can work in some circumstances and block the use of valuable and productive knowledge in others [3][5].

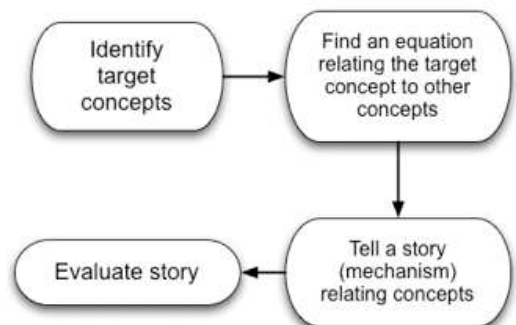

Fig. 7: Making meaning with mathematics – An epistemic game that can help students make sense of physics and blend their physical and mathematical knowledge [3][5].

The moves in recursive-plug-and-chug (Fig. 6) are plausible and useful moves in solving quantitative physics problems. The problem is with the moves that have been omitted. There is no move that says, "Evaluate whether the equation you have chosen is appropriate to explain the situation you are considering." A game that has these moves is making meaning with mathematics (Fig. 7). What is interesting is that students in algebra-based physics tend to play one game or the other and not blend them well.

In one example, we saw a student whose local goal was to estimate the volume of her dormitory room decide that the answer required was 1 m$^3$ – because that was the only volume she could find in the problem statement. She was playing recursive-plug-and-chug, and in this game, all information must come from an authoritative source; you are not allowed to reach into your life experiences for an answer. Unfortunately, the problem that was posed was an estimation problem – one in which students were explicitly expected to use their everyday knowledge in order to construct solutions. This student had categorized the problem differently from her instructor and (tacitly) chosen the wrong e-game to play. This mismatch of expectations between student and teacher is quite common and has been observed with upper division physics students as well [6].

## 4. What are the implications for our teaching?

From this analysis of the use of math in physics (and in science in general), we have learned a number of important results that have implications for our teaching. There's more to problem solving than learning "the facts" and "the rules." What expert physicists do in even simple problems is quite a bit more complex than it may appear to them and is not "just" what is learned (or not learned) in a math class. Helping students to learn to recognize what tools (games) are appropriate in what circumstances is critical.

Physics is an excellent place for scientists in many fields to learn to use mathematics in science but too much of an emphasis on algorithmic approaches can block students from learning other important parts of how to approach physics problem solving if those other parts are not taught. We need to improve our understanding of the cognitive processes involved in physics problem solving and find activities that help our students build knowledge into intuitions/understanding. Building manipulation skills is not enough.

### Acknowledgments

This work is supported in part by the National Science Foundation under Grants No. REC-008 7519 and DUE 05-24987. Any opinions, findings, and conclusions or recommendations expressed in this material are those of the author(s) and do not necessarily reflect the views of the National Science Foundation.